\documentclass[11pt]{article}
\usepackage{geometry}    
\usepackage{latexsym}
\usepackage{graphics}
\usepackage{graphicx}
\usepackage{epstopdf}
\usepackage{epsfig}
\usepackage{amssymb}
\usepackage{makeidx}
\usepackage{amsfonts}
\usepackage{amstext}
\usepackage{amsmath}
\usepackage{amsbsy}
\usepackage{wasysym}
\usepackage{mathrsfs}
\usepackage{authblk}
\DeclareMathOperator{\erfc}{erfc}

\newlength{\InitialBLS}
\author[a,*]{E. Segreto}
\author[b]{A.A. Machado}
\author[b]{L. Paulucci}
\author[c]{F. Marinho}
\author[d] {D. Galante}
\author[a]{S. Guedes}
\author[a]{A. Fauth}
\author[d] {V. Teixeira}
\author[a]{B. Gelli}
\author[a]{M. Reggiani-Guzzo}
\author[d] {W. Araujo}
\author[d] {C. Ambr\'osio}
\author[d] {M. Bissiano}
\author[a]{A.L. Lixandr\~ao Filho}

\affil[a]{Instituto de F\'isica Gleb Wataghin, Universidade Estadual de Campinas - Unicamp, Rua Sergio Buarque de Holanda, No 777, CEP 13083-859 Campinas, SP, Brazil}

\affil[b]{Universidade Federal do ABC (UFABC), Av. dos Estados, 5001, Santo Andr\'e, SP, 09210-170, Brazil}

\affil[c]{Universidade Federal de S\~ao Carlos, Rodovia Anhanguera, km 174, 13604-900, Araras, SP,
	Brazil}

\affil[d]{Laborat\'orio Nacional de Luz S\'incroton (LNLS), Rua Giuseppe M\'aximo Scolfaro, 10000 - Guar\'a, Campinas - SP, 13083-100, Brazil}

\title{Liquid Argon test of the ARAPUCA device}

\begin{document}
	\maketitle
	\let\oldthefootnote\thefootnote
	\renewcommand{\thefootnote}{\fnsymbol{footnote}}
	\footnotetext[1]{Corresponding author - E-mail: segreto@ifi.unicamp.br}
	\let\thefootnote\oldthefootnote
	
\begin{abstract}
	The ARAPUCA is a novel concept for liquid argon scintillation light detection which has been proposed for the photon detection system of the Deep Underground Neutrino Experiment. The test in liquid argon of one of the first ARAPUCA prototypes is presented in this work, where the working principle is experimentally demonstrated. The prototype has an acceptance window of 9 cm$^2$ and is read-out by a single SiPM with active area of 0.36 cm$^2$. Its global detection efficiency was estimated by exposing it to a $^{238}U$ $\alpha$ source and to cosmic rays and was found to be 1.15\% $\pm$ 0.15\%, in good agreement with the prediction of a detailed Monte Carlo simulation of the device. Several other ARAPUCA prototypes of bigger dimensions and read-out by arrays of SiPMs have been built and are actually under test. In particular 32 ARAPUCA cells have been installed inside the protoDUNE detector, which is being assembled at CERN and will be operated in the second half of 2018.
\end{abstract}	

\section{Introduction}
Neutrino oscillations physics actually represents one of the most powerful tools to investigate new physics beyond the Standard Model.

The Deep Underground Neutrino Experiment (DUNE) \cite{DUNE} is a leading-edge experiment aimed to address some of the open questions in this field, such as the measurement of the CP violating phase in 
the leptonic sector, the hierarchy of neutrino masses and the $\theta_{23}$ octant.

The Long-Baseline Neutrino facility (LBNF) will provide a high intensity, broad band neutrino beam which will originate from the Fermi 
National Accelerator Laboratory (Fermilab), Illinois, and will propagate for 1,300 km up to the DUNE far detector. It will be constituted by a liquid argon (LAr) time projection chamber (TPC) with active mass of 40 kt
and will be installed at the Sanford Underground Research Facility in South Dakota, located 1,300 km away. This is the ideal distance since it will give the highest sensitivity in measuring the CP violating phase and in determining the mass hierarchy \cite{DUNE2}.
DUNE also foresees the realization of a near detector located at the Fermilab to monitor the neutrino flux at production and reduce the systematics of the experiment.

The huge active mass of the far detector will also allow to develop a rich program of non accelerator physics, that includes the search for proton decay and the detection of supernova and atmospheric neutrinos.

The experimental technique of LArTPC takes advantage of the detection of LAr scintillation light for various purposes. LAr 
is known to be an abundant scintillator, emitting 40 photons per keV of deposited energy by minimum 
ionizing particles. Scintillation photons are emitted through the transition of the $Ar_2^*$ excimer 
to the ground dissociative state made of two separate argon atoms. The time evolution of the emission is 
described by the sum of two exponentially decaying distributions - one with a characteristic time of 6 
nsec and the other of 1300 nsec \cite{doke}. The relative abundance of the two components depends on the  particle which caused the excitation. The detection of the fast component is often used for triggering 
purposes and for determining the time of occurrence of the ionizing event in LAr ($t_0$ time), which is 
fundamental for position and calorimetric reconstructions. Scintillation light can also be used to 
perform calorimetric measurements and for particle identification. 
Photons are emitted in the Vacuum Ultra Violet (VUV) in a ten nm band centered around 127 nm \cite{doke}, where the vast majority of commercial photo-sensors is not sensitive. 
A convenient way to detect LAr scintillation photons is to convert their wavelength to the visible through wavelength shifting compounds and make them detectable to standard (cryogenic) photo-sensitive devices with glass or fused 
silica window.\\
The scintillation light detection system for a massive LArTPC, like the one foreseen for the DUNE far 
detector, is challenging since it needs to meet the discording requirements of a large coverage - given the internal dimensions of the detector - and of a good detection efficiency, at the 
level of percent - to meet the physics goals of the experiment - while keeping the costs at an acceptable level.

The ARAPUCA \cite{Machado2016} is a device proposed for the light detection system of the DUNE far detector. It is constituted by a light collector coupled to silicon photo-sensors (SiPM) used 
to read-out the collected photons. 
The collector consists of a box cavity with highly reflective internal surfaces from which photons can reflect back and forth until being detected by the SiPM array (or absorbed). The acceptance window of the box is made by a short-pass dichroic filter, which has the properties of being highly transparent to photons with wavelength below a given cut-off, while being highly reflective to photons with wavelength above the same cut-off.
VUV scintillation light is shifted one time outside the box, by a wavelength shifting film deposited on the external side of the dichroic filter to a wavelength below the cut-off to allow the photons to enter inside the collector box and a second time inside the box, to a wavelength above the cut-off to prevent photons from leaving the box.
The internal shifter can be deposited or directly on the filter or on the surfaces of the reflective box.
The shape of the ARAPUCA is constrained to be a flattened box by the mechanical requirements of the detector. In particular, it needs to be slided between the anodic wire plane assemblies and its total thickness can not exceed two centimeters.\\

The results of the test of one of the earliest prototypes of the ARAPUCA device in a LAr environment and exposed to an $\alpha$ source and to cosmic muons are presented. This is the demonstration of the experimental technique and represents the first LAr experiment in Latin America. Several other ARAPUCA prototypes have been fabricated up to now with bigger dimensions and an increased number of read-out SiPMs and are under test. In particular an array of 32 ARAPUCAs has been installed in the protoDUNE detector which is being assembled at CERN and which will be operated on a charged particle beam starting from late 2018. The dimensions of the acceptance window of the protoDUNE ARAPUCAs are 10 cm $\times$ 8 cm and an array of 12 or 6 SiPMs is installed inside each reflective box.    

\section{Experimental set-up}

\begin{figure}
	\begin{center}
		\includegraphics[width=6cm]{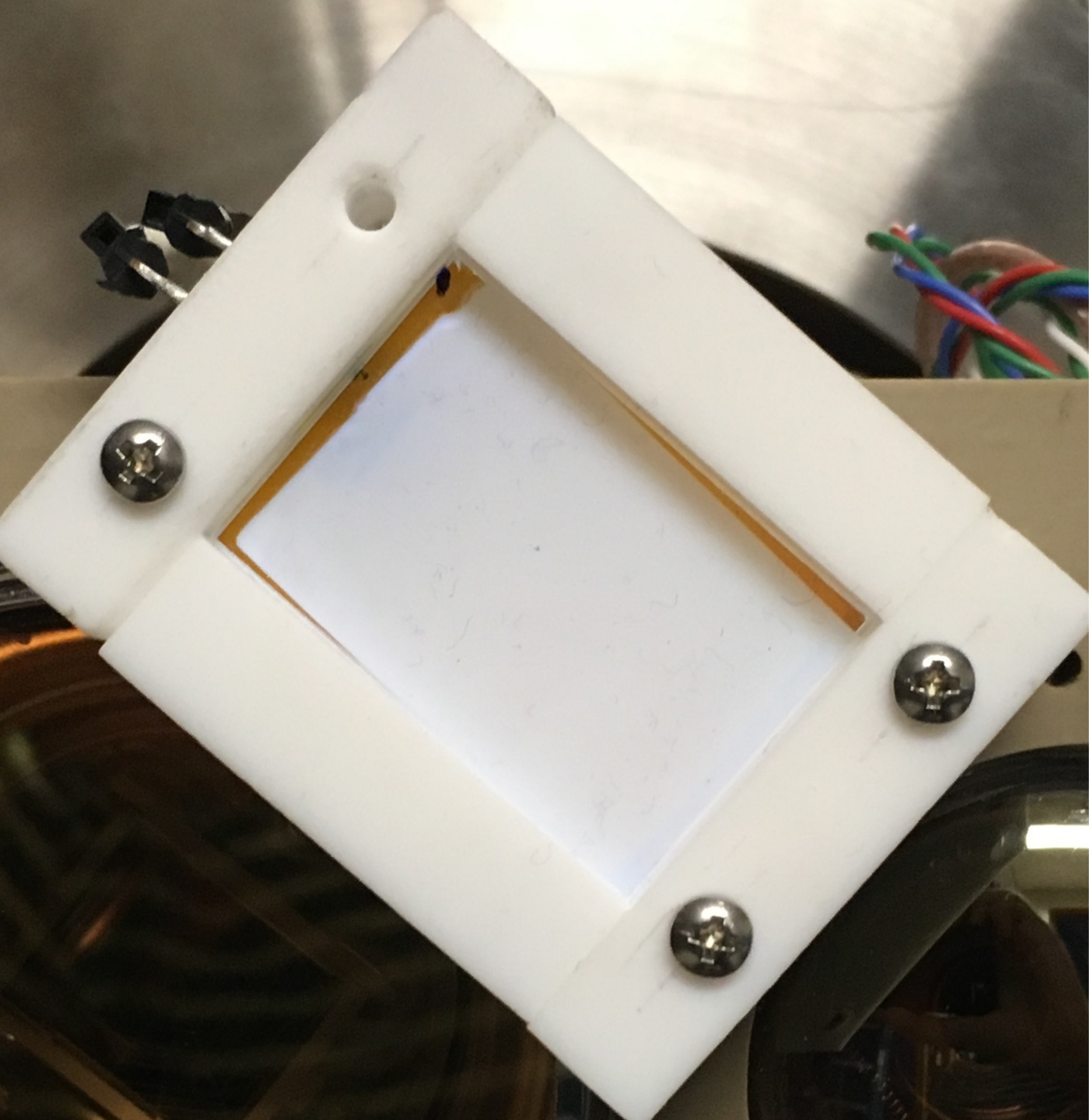}
		\includegraphics[width=8.25cm]{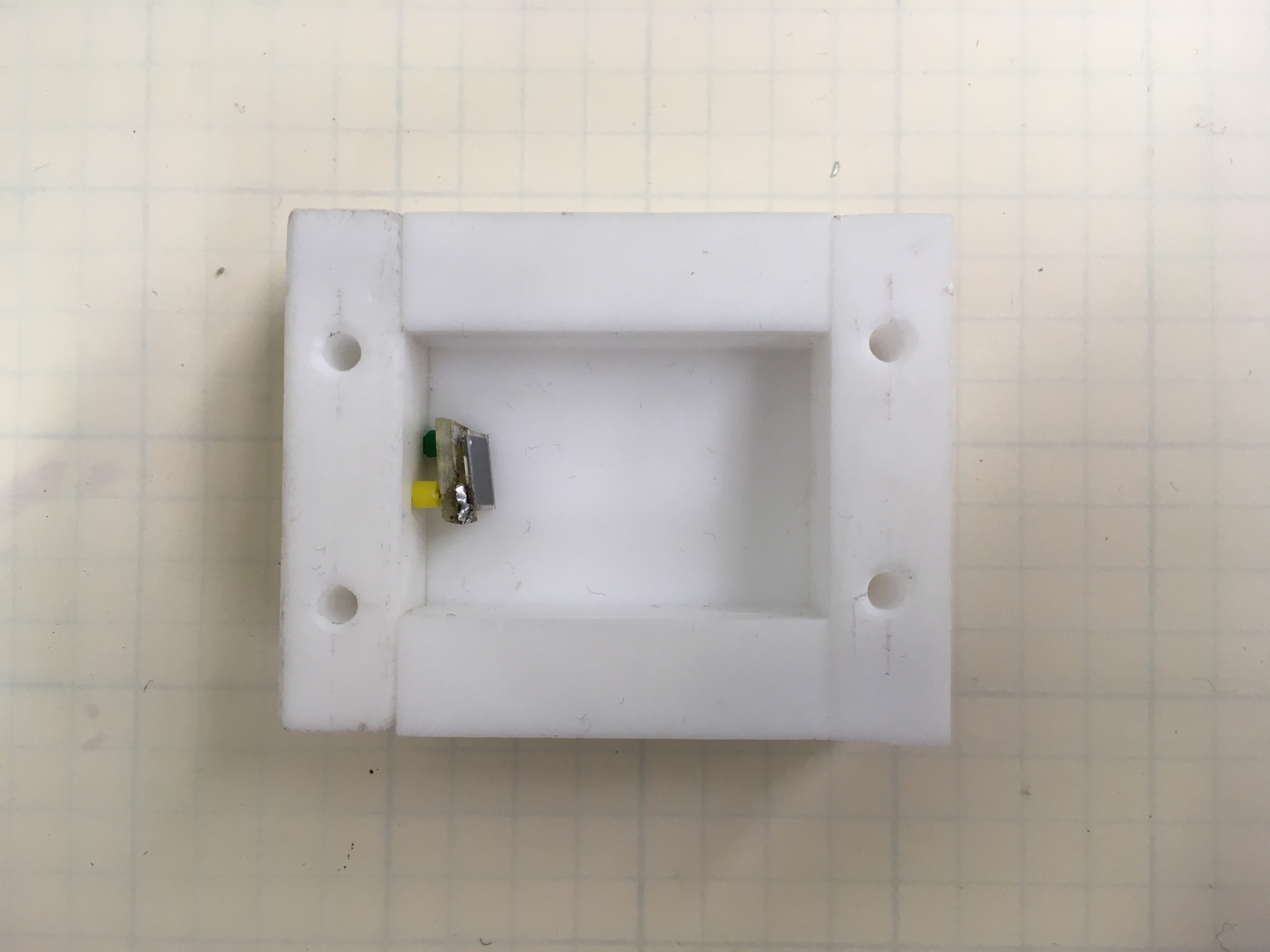}
		\caption{Left: the fully assembled ARAPUCA device. Right: a view of the inside of the reflective cavity with the installed SiPM.}
		\label{arapuca_box}
	\end{center}
\end{figure}

The ARAPUCA device which have been considered for this test is constituted by a PTFE box with internal dimensions of 3.6 cm $\times$ 2.5 cm $\times$ 0.6 cm, with an acceptance window made of a dichroic filter with dimensions of  3.6 cm $\times$ 2.5 cm  (2 mm thick) and cut-off at 400 nm.
The filter was coated on the external side with a film of p-Terphenyl (pTP) \cite{pTP} and on the internal side with a film of TetraPhenyl-Butadiene (TPB) \cite{TPB} \cite{TPB_ettore}. The pTP has an emission spectrum peaked around 350 nm, in the region where the dichroic filter is transparent while the TPB has its emission peaked around 430 nm where the filter turns to be reflective. The wavelength shifting films were produced by vacuum evaporation at the Central Mulitiusuario of the Federeal University of ABC (UFABC). The surface density of the pTP and TPB films was of 500 $\mu$g/cm$^2$ in both cases. A picture of the complete set-up is shown in figure \ref{arapuca_box}, left. One single SiPM mod. SensL MicroFC-60035-SMT \cite{cseries} was installed on one of the short lateral sides of  the inner box, as shown in figure \ref{arapuca_box}, right.

\subsection{Cryogenic set-up}
The cryogenic test of the ARAPUCA was performed at the facilities of the Toroidal Grating Monochromator (TGM) beamline of the Brazilian Synchrotron Light Laboratory (LNLS). 

The ARAPUCA was installed at the bottom of a vacuum tight stainless-steel cylindrical vessel with an internal volume of 2.3 liters ($\diameter$ $\simeq$ 10 cm - internal height $\simeq$ 30 cm). An $\alpha$ source was positioned at a distance of 4.0 cm from the ARAPUCA window, in correspondence of the center of its window. The source was an $^{238}$U-Al alloy in the form of a thin metallic circle with diameter of 5 mm and thickness of about 14 $\mu$m and it was screwed on a PTFE frame.\footnote{The frame was perforated for the 90\% of its surface to avoid that photons converted on the ARAPUCA window and not trapped could be reflected back and artificially increase the efficiency. The PTFE reflectivity to 128 nm light is negligible \cite{Kadkhoda}}\\  
Alpha particles are emitted with an energy of E$_{\alpha}$ = 4.267 MeV, but since the $^{238}$U atoms are embedded in a aluminum matrix, they emerge with a continuous residual energy spectrum with end point at  E$_{\alpha}$. A dedicate and independent test was performed in order to confirm that the parent nuclei emitting $\alpha $ particles is actually $^{238}$U.
The Poly Allyl Diglycol Carbonate (PADC) solid state $\alpha$ detector was used to characterize the source. The $\alpha$ particle interacts with the polymeric chains creating a modified path, usually called track. The PADC is chemically etched and etch pits with diameters of about 10 $\mu$m are produced around the alpha particle tracks, which can be observed under an optical microscope. Etch pit sizes and gray levels are correlated with the alpha particle energy \cite{Hadler} \cite{Santos} \cite{Soares}. The comparison of the etch pits sizes and gray levels produced by the source under investigation with those produced by known sources confirmed that it is compatible with $^{238}$U of almost infinite thickness\footnote{The range of 4.267 MeV $\alpha$ particles in aluminum is about 20$\mu$m.}.

\begin{figure}
	\begin{center}
		\includegraphics[width=0.7\textwidth]{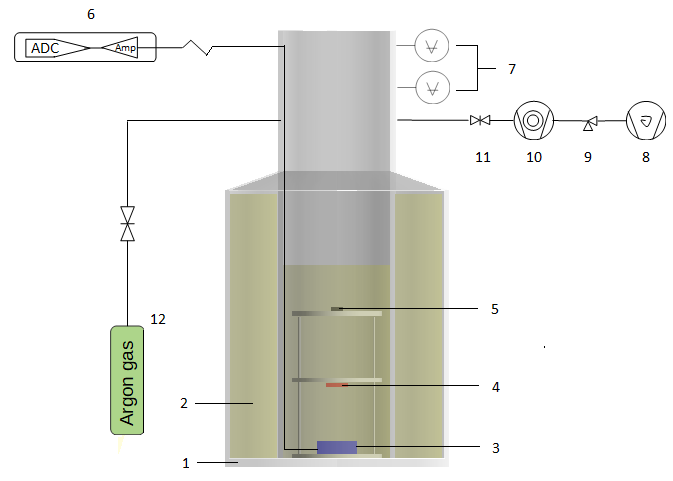}
		\caption{Schematic diagram of the vacuum chamber with gas system and the ARAPUCA assembly, inside a dewar with a liquid argon bath. The experimental setup is divided into cryogenic system (1. dewar and 2. liquid argon), ARAPUCA assembly (3. ARAPUCA, 4. radiation source, 5. level sensor, 6. electronic acquisition system), vacuum system (7. pressure sensors, 8. primary vacuum pump, 9. pre-vacuum valve, 10. turbomolecular pump) and gas system (11. vacuum gate valve and 12. argon gas cylinder).}
		\label{fig:setup}
	\end{center}
\end{figure}

In order to achieve the highest possible LAr purity inside the ARAPUCA's chamber, the stainless-steel cylinder assembly was pre-baked and vacuum pumped down to a pressure of 10$^{-6}$ mbar prior to filling it with pure argon (grade 6.0). The level of the liquid was monitored with a level meter
and was set to an height able to submerge completely bot  the $\alpha$ source and the ARAPUCA.
An open cryostat filled with commercial argon was used as a thermal bath for the system. A schematic representation of the experimental set-up is shown in figure \ref{fig:setup}.

The output signal of the SiPM was processed by a preamplifier specifically developed for this application (gain of 20 dB and 10 kHz to 500 MHz operating bandwidth) and then recorded by a fast Waveform Recorder (Acqiris, DP235 Dual-Channel PCI Digitizer Card) installed on a PC, where it was stored to be further processed. The signal waveforms passing a threshold set at a level corresponding to few photo-electrons were recorded with sampling time of 1 ns over a full record length of 10 $\mu$s, with the use of a LabVIEW-based interface. The SiPM was biased with a Keithley  2231-A-30-3 power supply.

\section{Data analysis and results}
The test was entirely dedicated to the measurement of the global detection efficiency of the ARAPUCA prototype for liquid argon scintillation light. It was started on November 6$^{th}$ 2016 and lasted for about 24 hours. Scintillation light was produced by $\alpha$ particles or cosmic muons. $\alpha$ particles provide an ideal way to measure the detection efficiency, since they can be safely considered as a point like light source given that their range in LAr is of the order of tens of $\mu$m. The number of VUV photons impinging on the acceptance window of the ARAPUCA can be calculated through simple geometrical considerations, while in the case of muons a Monte Carlo simulation is needed to extract the same information.\\

\subsection{Single Electron Response}

The SiPM was biased at an over-voltage of 3 Volt along the entire test.
The number of photons detected by the SiPM installed inside the ARAPUCA in coincidence with the ionizing events in LAr is proportional to the integral of the charge contained in each recorded waveform. The calibration factor which allows to convert the waveform integral into number of detected photons, or photo-electrons, is the Single Electron Response (SER) of the SiPM, which is the average value of the integral of the waveform produced by the detection of one single photon.

\begin{figure}
	\begin{center}
		\includegraphics[width=0.7\textwidth]{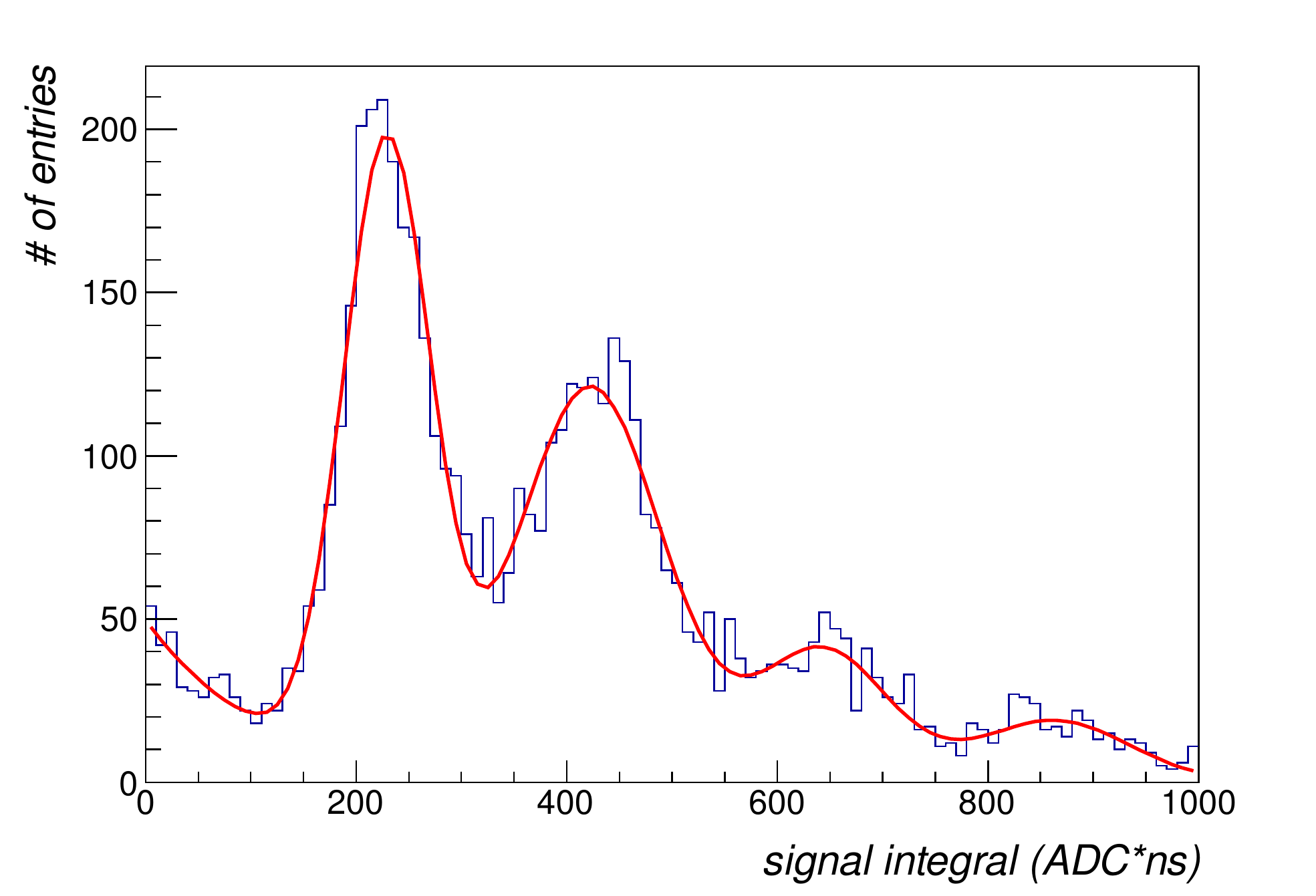}
		\caption{Single Electron Response of the SiPM. Red line represents a fit of the experimental distribution with a multi-Gaussian distribution.}
		\label{fig:SER}
	\end{center}
\end{figure}

Recorded waveforms have been individually analyzed. In order to determine the SER of the SiPM in LAr, single photon signals have been searched in the tail of the scintillation signals, starting from 5 $\mu$s after the onset of the signal itself. A hit find algorithm has been implemented, which searches for single, isolated peaks and 
integrates their charge over 500 ns. The histogram of the integrals  for a large number of isolated peaks is shown in figure \ref{fig:SER}. The first peak around zero is produced by white gaussian noise, 
the second one by one photon signals, while the others by multiple photons (3 more peaks 
are visible, due to two, three and four photons). The distribution has been fitted with a function made 
by five Gaussians: the first one accounts for the white noise, while the others for the physical light 
signals. The mean value and standard deviation of the Gaussian distribution related to the single 
photon have been left as free parameters, while the mean value and standard deviation of the Gaussians 
distributions of multi-photons peaks are given by the 
relations:
\begin{equation}
\bar{x}_{n} = n\times x_0 
\label{eq:mean_value}
\end{equation}

\begin{equation}
\sigma_{n} = \sqrt{n} \times \sigma_{0}
\label{eq:sigma}
\end{equation}

where $\bar{x}_{n}$ is the mean value of the peak relative to the $n^{th}$ photon, $x_{0}$ is the mean 
value of the one photon peak, $\sigma_{n}$ is the standard deviation of the peak relative to the 
$n^{th}$ photon and $\sigma_{0}$ is the standard deviation of the one photon gaussian distribution.
The normalization factor of each one of the peaks has been left as a free parameter.\\

The result of the fit is shown in figure \ref{fig:SER} with a red line. x$_0$ represents the SER value for the SiPM and is used for the spectral analysis.

\subsection{Alpha spectrum}
\label{subsec:alpha_spec}
The ARAPUCA was exposed to the scintillation light produced by the $\alpha$ source immersed in LAr. A statistically significant number of muons was detected too, since the set-up was installed on surface and hence exposed to the cosmic ray flux. Exploiting the pulse shape capabilities of LAr it is possible to analyze separately the $\alpha$ and $\mu$ samples and obtain two independent estimations of the ARAPUCA efficiency.\\ 

The time evolution of any LAr scintillation signal, L(t), is described as the sum of two exponentially decaying distributions with very different characteristic times:

\begin{equation}
L(t) = \frac{A_f}{\tau_f}e^{-t/\tau_f}+\frac{A_s}{\tau_s}e^{-t/\tau_s}
\label{eq:scintillation_pulse}
\end{equation}

the two components are usually referred to as the fast and slow LAr scintillation components with characteristic times $\tau_f$ and $\tau_s$, which take values of $\sim$ 6 ns and $\sim$ 1300 ns respectively \cite{nitrogen}. 
The relative abundance of the two components, A$_f$/A$_s$, strongly depends on the type of radiation which caused the ionization of the LAr and it ranges from 0.3 for electrons to 3.0 for neutrons \cite{pulse_shape}. 
This strong dependence is at the basis of the excellent discrimination power of LAr, which is widely used in direct Dark Matter detection experiments for background rejection purposes \cite{DarkSide} \cite{Deap}.\\

In order to discriminate $\alpha$ and $\mu$ events, a pulse shape parameter, F$_{prompt}$, is defined, that accounts for the fraction of fast light contained in the scintillation signal:
\begin{equation}
F_{prompt} = \frac{\int_{t_0}^{t_0+1500~ns}I(t)~dt}{\int_{t_0}^{t_0+10000~ ns}I(t)~dt}
\end{equation}

where I(t) is the intensity of the detected scintillation signal and t$_0$ is its start time. 
The integration time of the fast component is much longer than usual($\sim$ 100 ns \cite{TPB_ettore}) because of the characteristic discharge time of the SiPM at LAr temperature which is of the order of hundreds of nanoseconds.

\begin{figure}[t]
	\begin{center}
		\includegraphics[width= 7cm]{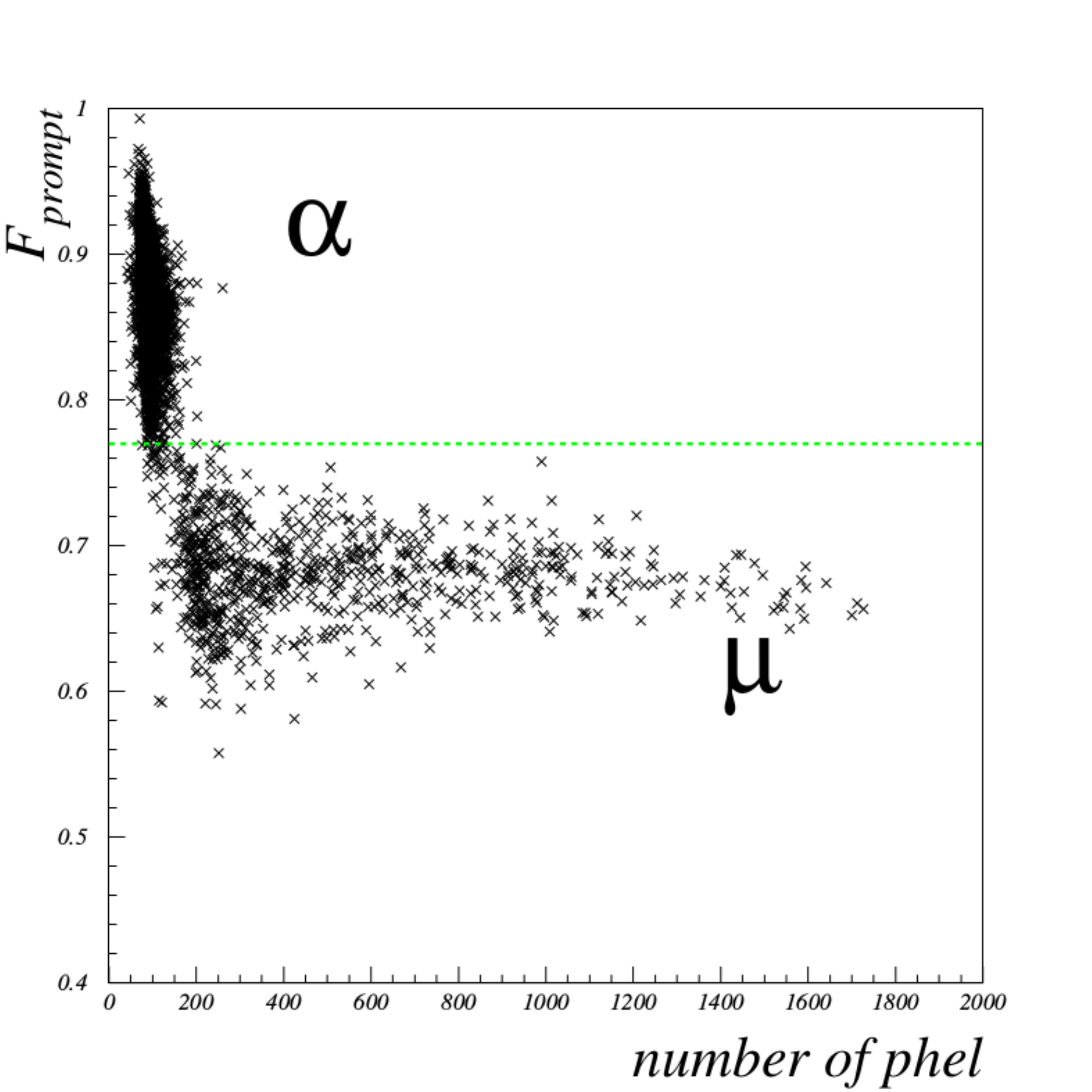}
		\includegraphics[width= 7cm]{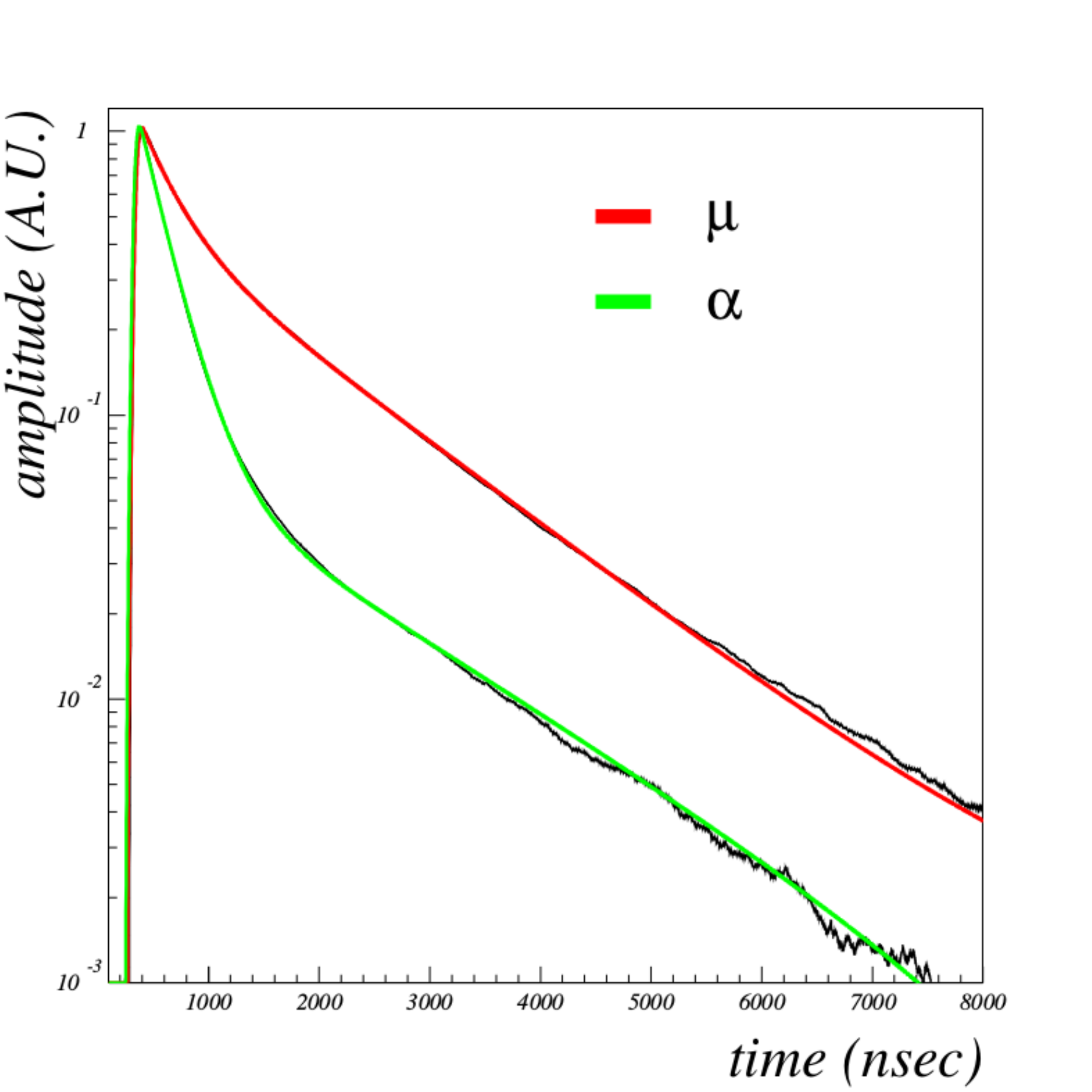}	
		\caption{Left: Distribution of F$_{prompt}$ for a sample of events. It is very clear the separation between $\alpha$ particles and muons. The green line represents the cut which has been set to discriminate between them and corresponds to F$_{prompt}$ = 0.77. Right: average waveforms for $\alpha$ particles (F$_{prompt}$>0.77) and muons (F$_{prompt}$<0.77). The green line is a fit of the $\alpha$ average waveform and the red one is a fit to the average muon waveform.}
		\label{fig:f_prompt}
	\end{center}
\end{figure} 

The scatter plot of F$_{prompt}$ as a function of the number of detected photo-electrons is shown in figure \ref{fig:f_prompt}, left. The $\alpha$ and $\mu$ populations are very clearly separated and a cut at F$_{prompt}$ = 0.77 is set to distinguish between them:
\begin{itemize}
	\item waveforms with a value of F$_{prompt}$ below 0.77 are assumed to be produced by muons
	\item  waveforms with a value of F$_{prompt}$ above 0.77 are assumed to be produced by alpha particles.
\end{itemize}	

The waveforms selected according to this criterion are used to calculate average waveforms for $\alpha$s and muons, which are shown in figure \ref{fig:f_prompt}, right. The two average waveforms have been fitted with a function which is the convolution of the sum of two decaying exponentials, to take into account the behavior of the scintillation pulse as described in equation \ref{eq:scintillation_pulse}, with a Gaussian function to take into account the statistical fluctuations of photons' production, propagation and detection and the electronic noise of the system. The result of the two fits is shown in figure \ref{fig:f_prompt}, right with a red and a green line for muons and $\alpha$s waveforms respectively. The main parameters coming from the fitting procedure are shown in table \ref{tab:wav_fit}.

\begin{table}[h]
	\begin{center}
		\caption{List of the most relevant parameters coming from the fit of the average waveforms for $\alpha$ particles and muons. The value of the fast scintillation component is not shown because it is affected by the decay time of the SiPM signal. Only statistical errors are quoted.}
		\label{tab:wav_fit}	
		\vspace{0.3cm}
		\begin{tabular}[h]{c|c|c}
			\hline\hline
			&$\mu$&$\alpha$\\
			\hline
			Abundance fast component & 23$\pm$1 \% & 71$\pm$2 \%\\
			Abundance slow component & 77$\pm$1 \% & 29$\pm$1 \%\\
			Decay time slow component &1470$\pm$3 nsec& 1810$\pm$30 nsec\\
			\hline\hline
		\end{tabular}	
	\end{center}		
\end{table}

\begin{figure}[t]
	\begin{center}
		\includegraphics[width= 10cm]{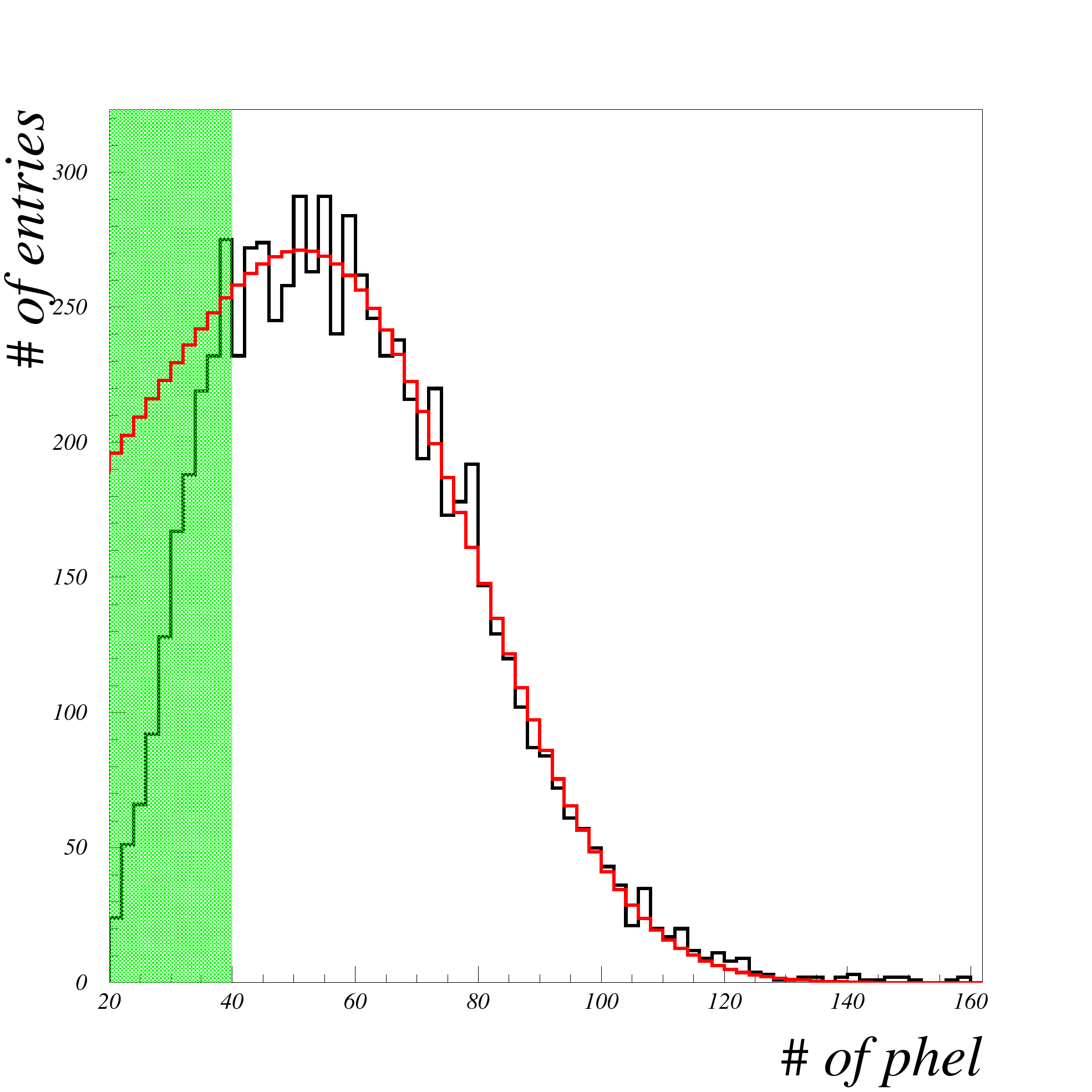}
		\caption{Spectrum of the total number of photo-electrons collected with the $\alpha$ sample (black line). The distribution is cut-off below 40 phel because of the trigger threshold. The red line is a fit of the experimental distribution with an analytical description of the spectrum (see text). In green the portion of the spectrum which is not considered in the fit due to the threshold effect.}
		\label{fig:alpha_spectrum}
	\end{center}
\end{figure} 

The decay time of the slow component of the average muon waveform is above 1400 nsec, which ensures that the liquid argon is clean, with a level of  contamination significantly below 1 pmm \cite{nitrogen} \cite{oxygen}. The ratio of the abundances of the fast to slow component is found to be of the order of 0.3 and 2.4 respectively for muons and $\alpha$ particles, consistent with what is reported in literature \cite{f_prompt_e_alpha}.
The presence of a third component (attributed to TPB fluorescence), which has been reported by many authors (see \cite{TPB_ettore} and references therein) can not be verified due to the long decay time of the SiPM signal ($\sim$ 300 nsec) which washes out this feature together with the short decay of the fast scintillation component.\\

The spectrum of the number of detected photo-electrons for the $\alpha$ sample is shown in figure \ref{fig:alpha_spectrum} with a black line. The spectrum is cut-off approximatively below 40 phel because of the trigger threshold which was set relatively high to reject the environmental electronic noise of the synchrotron light laboratory.

The energy spectrum of the $\alpha$ particles can be safely considered as produced by a mono-energetic source with infinite thickness and peak position at E$_{\alpha}$ = 4.267 MeV. It has been fitted with a convolution of an exponential low-energy tail with a Gaussian distribution:
\begin{equation}
f(E-\mu;\sigma,\tau)=\frac{A}{2\tau}\exp\Bigg(\frac{E-\mu}{\tau}+\frac{\sigma^2}{2\tau^2}\Bigg)\erfc\Bigg(\frac{1}{\sqrt{2}}\Bigg(\frac{E-\mu}{\sigma}+\frac{\sigma}{\tau}\Bigg)\Bigg)
\end{equation}
where $\mu$ is the energy of the peak (in photo-electrons), A is the peak area, $\sigma$ is the standard deviation of the Gaussian and $\tau$ is the tailing parameter. This has been shown to be  one of the most successful analytical models to represent the shape of a mono-energetic alpha peak \cite{pomme}. The result of the fit is shown with a red line in figure \ref{fig:alpha_spectrum} and it returns a value for the peak position of:
\begin{equation}
\mu = 76\pm 1~~~ photo-electrons
\end{equation}

The number of photons impinging on the ARAPUCA acceptance window, $N_{\gamma}^A$, can be estimated as:
\begin{equation}
N_{\gamma}^A = N_{\gamma}^{LAr}\times E_{\alpha} \times q_{\alpha} \times \Omega^A \simeq 6100~~\gamma
\label{eq:n_window}
\end{equation}
where N$_\gamma$$^{LAr}$ is the photon yield of LAr, E$_\alpha$ is the energy of the $\alpha$ particles emitted by the source, q$_\alpha$ is the quenching factor for scintillation light production of $\alpha$ particles in LAr and $\Omega^A$ is the solid angle subtended by the acceptance window of the ARAPUCA with respect to the position of the $\alpha$ source.
The parameters used for this estimation are listed in table \ref{tab:n_gamma}.

\begin{table}[h]
	\begin{center}
		\caption{Parameters used to evaluate the expression of equation \ref{eq:n_window}.}
		\label{tab:n_gamma}	
		\vspace{0.3cm}
		\begin{tabular}[h]{c|c|c}
			\hline\hline
			LAr photon yield @ 0-Field & N$_\gamma^{LAr}$ = 5.1 $\times$ 10$^4$ $\gamma$/MeV& \cite{doke}\\
			Energy of $^{238}$U $\alpha$ particles & E$_\alpha$ = 4.276 MeV & \\
			Quenching factor scintillation light for $\alpha$ & q$_\alpha$ = 0.71 & \cite{mei}\\
			Solid angle & $\Omega^A$ = 0.04 &\\
			\hline\hline
		\end{tabular}	
	\end{center}		
\end{table}

\begin{figure}[t]
	\begin{center}
		\includegraphics[width= 10cm]{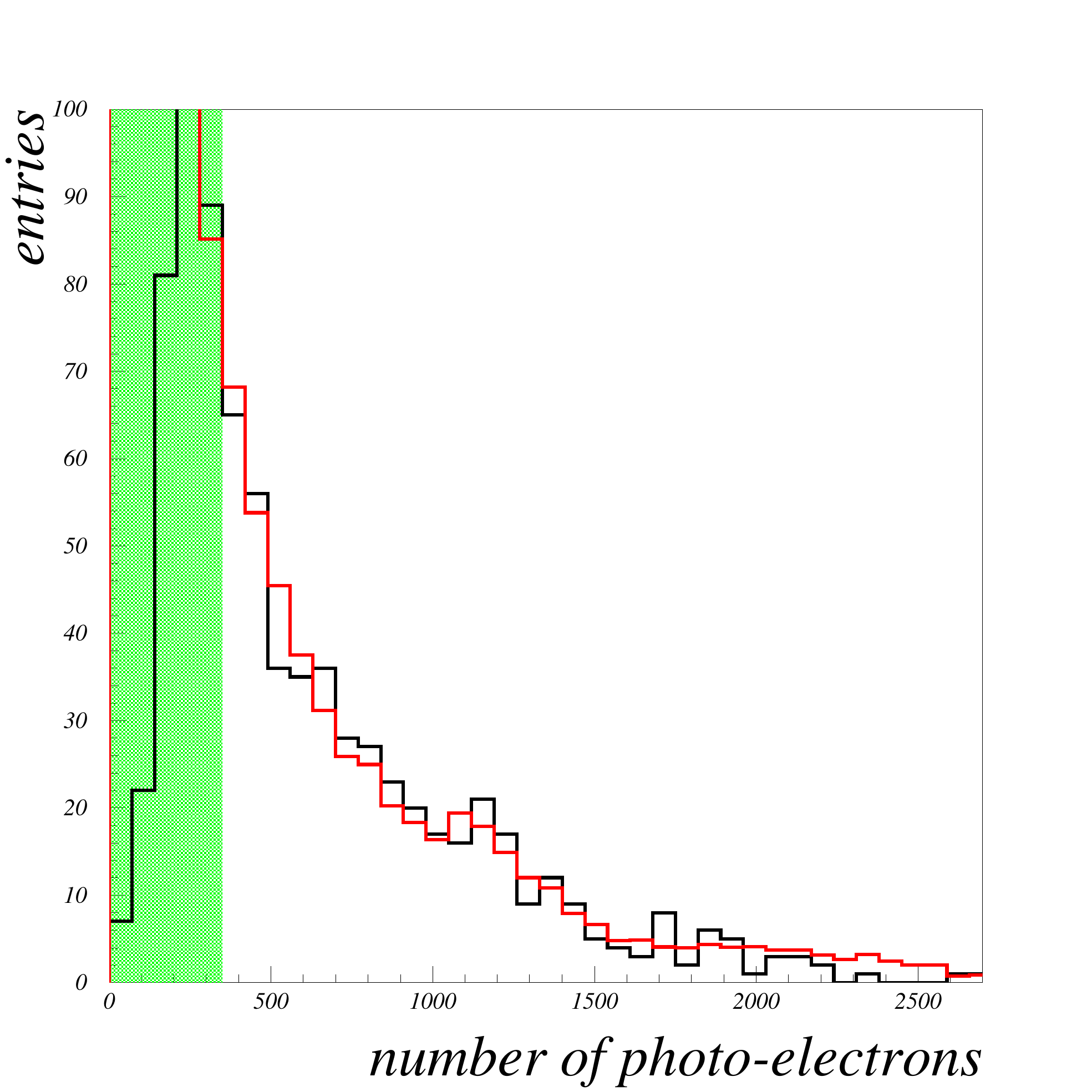}
		\caption{Spectrum of the total number of photo-electrons collected with the $\mu$ sample (black line). The distribution is cut-off below 350 phel because of the trigger threshold. The red line is a fit of the experimental distribution with the result of a Monte Carlo simulation. In green the portion of the spectrum which is not considered in the fit due to the threshold effect.}
		\label{fig:mu_spectrum}
	\end{center}
\end{figure} 

In order to complete the estimation of the efficiency of the ARAPUCA, three more effects need to be considered:
\begin{enumerate}
	\item the cross-talk probability for the SensL MicroFC-60035-SMT at 3 Volts of overvoltage is around 15\% \cite{sun}. The number of detected photo-electrons needs to be reduced by this same amount to take into account this effect;
	\item the reflectivity of the internal surfaces of the stainless steel chamber (to VUV and visible light) can increase the number of photons impinging on the ARAPUCA window with respect to what determined on simple geometrical basis. This effect was estimated analytically through \cite{paper_ettore} taking into account the geometry of the inner chamber and the reflectivity of the materials and it was found to be of the order of 8\% to 10\%;
	\item when the chamber was opened, it was discovered that the pTP and TPB films deposited on the ARAPUCA window were slightly damaged. About 10\% to 15\% of the film detached from the dichroic filter\footnote{This problem has been deeply investigated in the framework of the ARAPUCA R\&D program and it is attributed to the cleaning process which the filter undergoes before being coated. The procedures developed after the test described in this work allow to obtain very stable films which do not suffer any damage when immersed in LAr}. This typically happens during the cool-down of the system and has the net effect of reducing the number of photons which are wavelength shifted and thus can enter the ARAPUCA box. This effect compensates the increase of the number of photons impinging on the acceptance window due to the reflections inside the chamber. A total systematic error of 15\% has been added to the result to take into account both these effects.              
\end{enumerate}

The total ARAPUCA efficiency, $\epsilon^{A}_{\alpha}$ estimated through the $\alpha$ particle sample results to be:

\begin{equation}
\epsilon^{A}_{\alpha} = 1.1\% \pm 0.2\%
\label{eq:epsilon_alpha}
\end{equation} 

\subsection{Muon spectrum}
The spectrum of total number of photo-electrons detected for the $\mu$ sample is shown in figure \ref{fig:mu_spectrum} with a black line. A simple Monte Carlo (MC) simulation was realized in order to extract a global detection efficiency from this spectrum too and compare it with what obtained with the $\alpha$ sample. 
Muons' directions are extracted from a cos$^2$ $\theta$ distribution and a flat $\phi$ distribution, where $\theta$ and $\phi$ are the polar and azimuthal angles respectively. Muons which intercepts the stainless steel vessel containing the ARAPUCA are propagated and scintillation light is produced along the track. The number of photons emitted per centimeter of track, N$_\gamma$, is given by:
\begin{equation}
N_{\gamma} = N_{\gamma}^{LAr}\times q_e \times \frac{dE}{dx} \simeq 104\times10^3~~\gamma/ cm
\label{eq:N_gamma_muons}
\end{equation}
The values of the parameters used to evaluate equation \ref{eq:N_gamma_muons} are listed in table \ref{tab:N_gamma_muons}.
\begin{table}[h]
	\begin{center}
		\caption{Parameters used to evaluate the expression of equation \ref{eq:N_gamma_muons}.}
		\label{tab:N_gamma_muons}	
		\vspace{0.3cm}
		\begin{tabular}[h]{c|c|c}
			\hline\hline
			LAr photon yield @ 0-Field & N$_\gamma^{LAr}$ = 5.1 $\times$ 10$^4$ $\gamma$/MeV& \cite{doke}\\
			Quenching factor scintillation light for electrons & q$_e$ = 0.78 & \cite{quenching_electrons}\\
			Energy loss per cm of track in LAr for 4 GeV muons & dE/dx = 2.61 MeV/cm & \cite{muons} \\
			\hline\hline
		\end{tabular}	
	\end{center}		
\end{table}

It has been considered that the mean energy of muons at the ground is $\sim$ 4 GeV \cite{PDG} and it has been assumed that the quenching factor of the scintillation light measured for relativistic electrons holds also for nearly minimum ionizing muons, since the Linear Energy Transfers (LET) in the two cases are comparable (low LET values $\sim$ 1.6 MeV cm$^2$/g).\\   
Photons are emitted isotropically and a check is performed if they hit the ARAPUCA acceptance window or not. Secondary particles production, reflections and scattering processes of the photons are not considered.\\
The MC spectrum of the number of photons hitting the ARAPUCA window is fitted on the spectrum of detected photo-electrons, where the scale and normalization factors are left as free parameters. The result of the fit is shown in figure \ref{fig:mu_spectrum} with a red line.
The scale factor between the number of detected photo-electrons and the number of photons impinging on the ARAPUCA window gives the total detection efficiency of the device.\\
The fit result is corrected by the cross-talk of the SiPM and the same considerations on reflections and wavelength shifting films integrity are done, as described in section \ref{subsec:alpha_spec}.\\
The global detection efficiency coming from the muon sample results to be:
\begin{equation}
\epsilon^{A}_{\mu} = 1.2\% \pm 0.2\%
\label{eq:epsilon_muon}
\end{equation}

The two measurements are in good agreement and their average value, together with the propagated error, represents the best estimative of the ARAPUCA global efficiency with the actual set of data:

\begin{equation}
\epsilon^{A} = 1.15\% \pm 0.15\%
\label{eq:epsilon_average}
\end{equation}

In order to be able to compare the performances of the ARAPUCA in detecting photons with other devices it is interesting to define the equivalent surface of the device, $\bar{S}$, as the product of the area of its acceptance window, A, and of its efficiency $\epsilon$:
\begin{equation}
\bar{S} = \epsilon\times A
\label{eq:equivalent_surface} 
\end{equation}

The equivalent surface is proportional to the number of photons the device can detect when exposed to an uniform flux of light and is the ideal quantity to compare different devices. In particular it can be used to calculate the gain, G, of the ARAPUCA with respect to the single SiPM which is installed inside it:
\begin{equation}
G=\frac{\bar{S}_{ARAPUCA}}{\bar{S}_{SiPM}^{TPB}}
\end{equation}

that is the ratio between the equivalent surface of the ARAPUCA divided by the equivalent surface of the SiPM installed inside the box, assuming that it is coated with a film of TPB to make it sensitive to VUV light.\\
The equivalent surface of the coated SiPM can be estimated considering that its active surface is of 0.36 cm$^2$, its detection efficiency at 3 Volt of overvoltage and averaged on the spectrum of TPB is of the order of 0.3 \cite{cseries}, the conversion efficiency of TPB for 127 nm light is around 0.5 \cite{benson} and half of the shifted photons are lost because they go in the opposite direction with respect to the SiPM window.
The gain for this ARAPUCA prototype results to be:
\begin{equation}
G=\frac{0.013\times 9 cm^2}{0.3 \times 0.36~cm^2 \times 0.25} \simeq 3.8
\end{equation}

That means that the use of the ARAPUCA amplifies the equivalent surface of the active device which is installed inside it by four times.

\section{Monte Carlo Simulation}
A simulation of the complete ARAPUCA prototype presented in this work was developed using the Geant4 toolkit \cite{Agostinelli2003, Allison2016}.  The optical properties of the Teflon used to build the box were implemented along with the ones for the three layers composing the acceptance window of the device (pTP, dichroic filter and TPB in sequence). A detailed description of an earlier version of this simulation can be found in \cite{Marinho2018}. The main improvements of the current version consist in considering the refraction index of liquid argon as a function of wavelength \cite{nLAr}, the absorption coefficient of pTP and TPB \cite{pTP}, the inclusion of the glass substrate for the filter and accounting for the angular dependence of the transmissivity/reflectivity of the dichroic filter according to the data provided by the manufacturer \cite{asahi}. 

\begin{figure}[t]
	\begin{center}
		\includegraphics[width=0.9\textwidth]{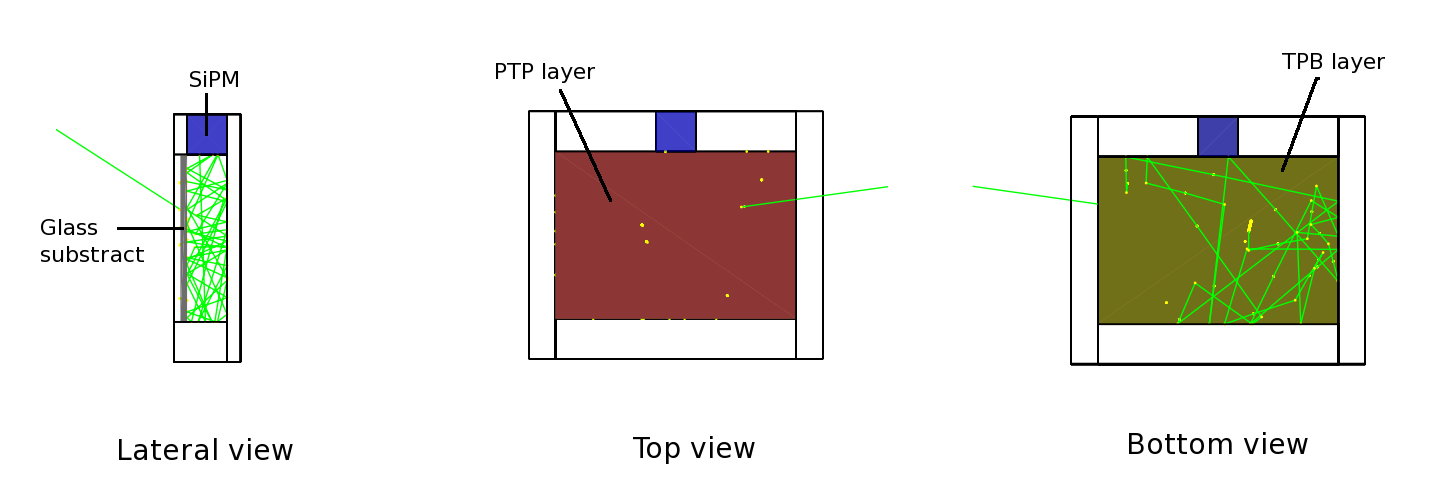}
		\caption{Lateral, top and bottom views of the simulated ARAPUCA device. The SiPM is located on one of the longer laterals (blue) and the box is made of highly reflective Teflon. The top view allows the visualization of the pTP layer (red) deposited on top of the fused silica substrate of the dichroic filter (gray) while the bottom view shows the TPB deposition (yellow). Photons are shown as (green) lines (color online).}
		\label{simulation}
	\end{center}
\end{figure}

A view of the simulated device is seen in figure \ref{simulation}. The dimensions and geometry used were chosen to reproduce the prototype tested. It has internal dimensions of 36 mm x 25 mm x 6 mm (length $\times$ width $\times$ height) and a Teflon reflectivity of 95\% for visible light. The SiPM has an active surface of 6 x 6 mm$^2$ and a quantum efficiency dependent on the photon's wavelength was associated (Sensl-C with an efficiency average of 30\%). Roughly half of the incident photons are emitted back when converted by the pTP while the remaining are effectively trapped inside the ARAPUCA. Some reach the SiPM after a few reflections (typically 2-3) while the remaining can be either detected after many reflections or be absorbed by the box internal surfaces. The simulation provided a detection efficiency estimate of 1.5\% $\pm$ 0.3\% for isotropic photons incident on the outside of the optical window. This result is in good agreement with the estimation presented in the previous section. The main sources of uncertainties are the parameters characterizing the shifters, in particular their absolute conversion efficiencies.

\section{Conclusions}
The ARAPUCA is a very new concept for the detection of LAr scintillation light which has been proposed for the photon detection system of the DUNE far detector. Its optical window, based on the use of a dichroic filter combined with two different wavelength shifting compounds, allows photons to enter the device but not to exit from it. Light results to be trapped inside a box with highly reflective internal surfaces and after few bunches can be eventually detected by the active sensor (SiPM) installed inside it. One of the first prototypes of this device has been tested in a LAr environment and has been exposed to an $^{238}$U $\alpha$ source and to cosmic rays to estimate its detection efficiency. The latter was estimated to be 1.15\% $\pm$ 0.15\%, in good agreement with the prediction of a detailed MC simulation of 1.5\% $\pm$ 0.3\%.

\section{Acknowledgments}
This work is funded by FAPESP (Funda\c{c}\~ao de Amparo \`a Pesquisa do Estado de S\~ao Paulo) under the projects 2016/01106-5, 2017/13942-5 and supported by CNPq (Conselho Nacional de Desenvolvimento Cient\'ifico e Tecnol\'ogico).

\end{document}